%% file: hal.tex
\documentclass[9.5pt,journal,final,finalsubmission,twocolumn]{IEEEtran}

\usepackage{listings}
\newtheorem{algorithm}{Algorithm}[section]

\def\fma{\mbox{\scriptsize \sf fma}}
\def\ulp{\mbox{ulp}}

\def\fractheo{0.44}

\newtheorem{theorem}{Theorem}%

\usepackage{color,url,cite,amstext,amssymb,amstext}
\DeclareFontFamily{U}{manual}{}
\DeclareFontShape{U}{manual}{m}{n}{ <->  manfnt }{}
\newcommand{\manfntsymbol}[1]{%
    {\fontencoding{U}\fontfamily{manual}\selectfont\symbol{#1}}}
\newcommand{\dbend}{\manfntsymbol{127}}%

\def\RN{\mathop{\circ}\nolimits}

\title{%
  Formally Verified Argument Reduction \\
  with a Fused-Multiply-Add
}%
\author{%
  Sylvie Boldo, Marc Daumas and Ren-Cang Li%
\thanks{%
  S.    Boldo ({\tt  sylvie.boldo@inria.fr}) is with the INRIA Futurs.}
\thanks{%
  M.    Daumas ({\tt marc.daumas@lirmm.fr}) is with the LIRMM, CNRS, UM2 and ELIAUS, UPVD. Supported in part
  by the PICS 2533 of the CNRS.}
\thanks{%
  R.-C. Li ({\tt rcli@uta.edu})  is  with the  Department
  of Mathematics, the University of Texas at Arlington,
        P.O. Box 19408,
        Arlington, TX 76019-0408. Supported in part
  by the National Science Foundation under Grant
        No. DMS-0510664 and DMS-0702335 and by the Region Languedoc Roussillon of France.}
}

\date{%
       February 2007}

\newcommand{\incb}[2]{\usepackage[#2]{#1}}

 \incb{graphicx}{dvips}
 \incb{epsfig}{dvips}

\newenvironment{pfigure}{\begin{figure*}}{\end{figure*}}
\def\figdir{.}

\begin{document}

\maketitle
\thispagestyle{empty}

\begin{abstract}
  Cody \& Waite argument reduction technique works perfectly for
  reasonably large arguments but as the input grows there are no bit
  left to approximate the constant with enough accuracy. Under mild
  assumptions, we show that the result computed with a
  fused-multiply-add provides a fully accurate result for many
  possible values of the input with a constant almost accurate to the
  full working precision. We also present an algorithm for a fully
  accurate second  reduction step to reach double full accuracy (all the significand bits of two numbers are significant) even in the worst cases of argument reduction. Our work recalls the common
  algorithms and presents proofs of correctness.  All the proofs are
  formally verified using the Coq automatic proof checker.
\end{abstract}

\begin{keywords}
Argument reduction, {\tt fma}, formal proof, Coq.
\end{keywords}

\section{Introduction}\label{sec:intro}

Methods that compute elementary functions on a large domain rely on
efficient argument reduction techniques.  The idea is to reduce an
argument $x$ to $u$ that falls into a small interval to allow
efficient approximations \cite{Mar2K,BDKMR05,Mul06,li:04}. A commonly used
argument reduction
technique~\cite{CodWai80,Mar90,StoTan99,Mar2K} begins with
one positive FPN (floating point number) $C_1$ to approximate a number $C>0$ (usually
irrational but not necessarily). Examples include $C=\pi/2$ or $\pi$
or $2\pi$ for trigonometric functions $\sin x$ and $\cos x$, and
$C=\ln 2$ for exponential function $e^x$.

Let $x$ be a given argument, a FPN.  The argument reduction starts by
extracting $\chi$ as defined by
\begin{eqnarray*}
x/C_1   & =       & \framebox{\parbox{2cm}{\centerline{$\chi$\vphantom{$k2^{-N}b$}}}}\,
                    \framebox{\parbox{1.5cm}{\centerline{$\varsigma$\vphantom{$k2^{-N}b$}}}}\,.
\end{eqnarray*}
Then it computes a reduced argument $x - \chi C_1$. The result is exactly a FPN as it is defined by an IEEE-754
standard remainder operation.
But division is a costly operation that is
avoided as much as possible. Some authors, see for example \cite{StoTan99,Mar2K,Mul06}, and
\begin{center}
  {\small \url{http://www.intel.com/software/products/opensource/libraries/num.htm}},
\end{center}
introduce another FPN $R$ that approximates $1/C$ and the argument
reduction replaces the division by a multiplication so that
\begin{eqnarray}
x\cdot\frac 1C   & \approx & xR   \nonumber \\
        & =       & z+s \nonumber \\
        & =       & \framebox{\parbox{2cm}{\centerline{$k2^{-N}$\vphantom{$k2^{-N}b$}}}}\,
                    \framebox{\parbox{1.5cm}{\centerline{$s$\vphantom{$k2^{-N}b$}}}}\,,
\label{eq:z-extract}
\end{eqnarray}
where $k$ is an integer used to reference a table of size $2^N$.
This replacement is computational efficient  if $u$
\begin{equation}\label{eq:u-1}
u =  x - z C_1
\end{equation}
is a FPN \cite{LiBolDau03}.

Sometimes the computed value of $u$ is not sufficiently accurate, for example if $u$ is near
a multiple of $C$, the loss of accuracy due to the approximation
$C_1 \approx C$ may  prevail. A better approximation to $C$
is necessary to obtain  a fully accurate reduced
argument. If this is the case we use  $C_2$, another FPN, roughly containing
the next many bits in the significand of $C$ so that the unevaluated
$C_1 + C_2 \approx C$ much better than $C_1$ alone.  When equation (\ref{eq:u-1}) does not introduce any rounding error,
the new reduced argument is not $u$ but $v$ computed by
\begin{eqnarray}\label{eq:u-updated}
v & \approx & u -z C_2.
\end{eqnarray}
To increase once again the accuracy, the error of (\ref{eq:u-updated})
need to be computed (see Section~\ref{sec:sndstep}), too to obtain
$v_1$ and $v_2$ exactly satisfying
\begin{eqnarray}\label{eq:u-updated}
v_1 + v_2 & = & u -z C_2.
\end{eqnarray}
The last step creates a combined reduced argument stored in the
unevaluated sum $v_1 + w$ with $2p$ significant bits
\begin{eqnarray}\label{eq:u-updated}
w & \approx & v_2 -z C_3
\end{eqnarray}
Whether $v_1$ (or $v_1$ and $w$) is accurate enough for computing the
elementary function in question is subject to further error analysis
on a function-by-function basis \cite{Kah83}. But this is out of the
scope of this paper.

The Cody \& Waite technique \cite{CodWai80} is presented in
Figures~\ref{fig/cody} and
\ref{fig/cody2}, where $\RN(a)$ denotes the FPN obtained from
rounding $a$ in the round-to-nearest mode. Those are examples when no {\tt fma} is used. The
sizes of the rectangles represent the precision (length of the
significand) of each FPN and their positions indicate magnitude, except for $z$ and $C_1$
whose respective layouts are only for showing lengths of significands. The light gray
represents the cancellations: the zero bits due to the fact that
$|x-\circ(z \times C_1)| \ll |x|$. The dark grey represents the
round-off error: the bits that may be wrong due to previous
rounding(s).

Figure~\ref{fig/cody} presents the ideal behavior.
Figure~\ref{fig/cody2} presents the behavior when the significand of
$z$ is longer. Then, fewer bits are available to store the significand
of $C_1$ in order for $z C_1$ to be  stored exactly. The consequence is a
tremendous loss of precision in the final result: as $C_1$ must be stored in
fewer bits, the cancellation in the computation of $x -z C_1$ is
smaller and the final result may be inaccurate.

\begin{figure*}[t]
\begin{minipage}{0.48\linewidth}
\centerline{\scalebox{.7}{\input{\figdir/cody.pstex_t}}}
\caption{Reduction technique works for $z$ sufficiently small.}
\label{fig/cody}
\end{minipage}
\hfill
\begin{minipage}{0.48\linewidth}
\centerline{\scalebox{.7}{\input{\figdir/cody2.pstex_t}}}
\caption{Cody \ Waite technique fails as $z$ grows, i.e. $u$ is not
  accurate enough.}
\label{fig/cody2}
\end{minipage}
\end{figure*}

\begin{figure*}
\centerline{\scalebox{.7}{\input{\figdir/enh.pstex_t}}}
\caption{Argument reduction with exact cancellation in a fused-multiply-add.}
\label{fig/enh}
\end{figure*}
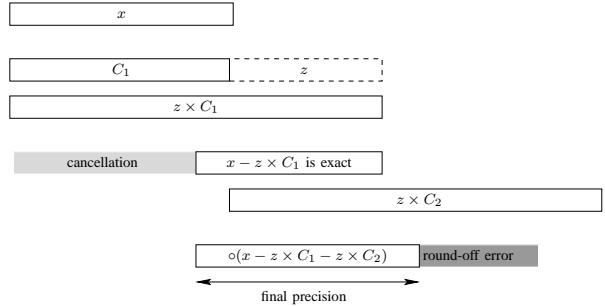

We want to take advantage of the fused-multiply-add ({\tt fma})
instructions.  Some machines have hardware support for it, such as
machines with HP/Intel$^\text{\textregistered}$
Itanium$^\text{\textregistered}$ Microprocessors~\cite{Mar2K} and IBM
PowerPC Microprocessors, and this instruction will also be added to
the revision of the IEEE-754 standard.  The current draft can be found
at
\begin{center}
   {\small \url{http://www.validlab.com/754R/}}.
\end{center}
It is obvious that some bits of $x$ and $z C_1$ will cancel each other
as $z$ is computed such that $x \approx z C_1$, but it is not clear
how many of them will and under what condition(s).  Consequently if
accuracy calls for $x - z C_1$ to be calculated exactly (or to more
than $p$ bits in the significand), how do we get these bits
efficiently?  This question is especially critical if the working
precision is the highest available on the underlying computing
platform.

In this paper, we will devise easily met conditions so that $x - z C_1$
can be represented exactly by a FPN, and thus it can be computed by one
instruction of the {\tt fma} type without error. This technique is
presented in Figure~\ref{fig/enh}. The understanding is the same as in
Figures \ref{fig/cody} and \ref{fig/cody2}. The cancellation is
greater as $C_1$ can be more precise.  The idea is that  the rounding
in $z C_1$ is avoided thanks to the {\tt fma}: $z C_1$, a
$2p$-bit FPN, is virtually computed with full precision and then
subtracted from $x$. This subtraction is proved to be exact as $x
\approx z C_1$. The fact of $x - z C_1$ being a FPN is used by the
library of \cite{StoTan99,Mar2K} with no formal justification
until \cite{LiBolDau03}.

The motivations of this work are similar to those presented in
\cite{LiBolDau03} and Section~\ref{sec:exact-sub} recalls briefly some
useful prior-art from the authors \cite{DauRidThe01,LiBolDau03}.
However, the rest of the paper presents entirely new results.  The
theorems and their proofs are different from the ones presented in
\cite{LiBolDau03}.  The changes are necessary to facilitate
verification with an automatic proof checker. Moreover, the results
have been improved, and are simpler to grasp and new results have been
added thanks to this simplification and to a better understanding of
the FPNs relationships due to the formal proof.

In a floating-point pen-and-paper proof, it is difficult to be
absolutely sure that no special case is forgotten, no inequality
is erroneous, and no implicit hypothesis is assumed, etc. All the
proofs presented in this paper are verified using our specification of
generic floating point arithmetic \cite{DauRidThe01} and Coq proof
assistant \cite{BC04}. This approach has already been proven successful in
hardware or software applications
\cite{CarMin95,Rus98,Har97a,Har2K}. The drawback is a long and
tiresome argumentation versus the proof checker that will ascertain
each step of the demonstration. The corresponding scripts of proofs
are available online at
\begin{center}
   {\small \url{http://www.netlib.org/fp/fp2.tgz}}.
\end{center}
We indicate for each theorem its Coq name.  The developments presented
here are located in the {\tt FArgReduct[2,3,4].v} files.

The rest of this paper is organized as follows.
Section~\ref{sec:exact-sub} recalls theorems on the number of
cancelled bits of two close FPNs (extensions of Sterbenz's
theorem~\cite{Ste74}). In Section~\ref{sec:alg}, we present the Coq
verified theorem about the correctness of the algorithm that produces
$z$ in (\ref{eq:z-extract}) and that satisfies the conditions of
the following theorems. The demonstration of the main result, 
{\it i.e.} the correctness of the first reduction step, is then 
described in Section~\ref{sec:proof}. In Section~\ref{sec:sndstep}, we give new
algorithms and results about a very accurate second step for the
argument reduction. Section~\ref{sec:concl} concludes the work of this paper.

\medskip\noindent
{\bf Notation.}  Throughout, $\ominus$ denotes the floating point
subtraction.  $\{{\cal X}\}_{\fma}$ denotes the result by an
instruction of the fused-multiply-add type, i.e., the exact $\pm a \pm
b \times c$ after only one rounding. FPNs use $p$ digits, hidden bit (if any) counted, in the
significand or otherwise explicitly stated.  We denote $\RN(a)$ the FPN obtained from
rounding $a$ in the round-to-nearest mode with $p$ digits and $\RN_m(a)$ if we round to $m$ digits instead of $p$. 
We denote by $\ulp(\cdot)$
the unit in the last place of a $p$-digit FPN and $\ulp^{\circ
  2}(\cdot) = \ulp(\ulp(\cdot))$.  
The smallest (subnormal)
positive FPN is denoted by $\lambda$.

\section{Exact Subtraction Theorems}\label{sec:exact-sub}
\setcounter{equation}{0}

These theorems will be used in Section~\ref{sec:proof} to guarantee
that there will be enough cancellation in $x-z  C_1$ so that it can be computed exactly
by one {\tt fma} type instruction, or equivalently, to assure
$x-zC_1$ fits into one FPN.

A well-known property \cite{Ste74,Gol91} of the floating point
subtraction is the following.

\medskip

\fbox{
\begin{minipage}{\fractheo\textwidth}
\begin{theorem}[{{\tt Sterbenz {\it in} Fprop.v}}]\label{thm:exact-sub-classical}
  Let $x$ and $y$ be FPNs. If  $y/2 \le x \le 2\ y,$
  then $x-y$ is a FPN.
  This is valid with any integer radix $\beta \ge 2$ and any
  precision $p \ge 2$.

\end{theorem}
\end{minipage}
}

\medskip

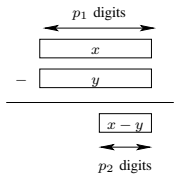
\begin{figure}[hb]
\centerline{\scalebox{.7}{\input{\figdir/Sterbenz.pstex_t}}}
\caption{Extension of Sterbenz's theorem.}
\label{fig/Sterbenz}
\end{figure}

We extend Sterbenz's theorem to fit the use of a fused-multiply-add
that may create a higher precision virtual number whose leading digits are canceled to the
working precision as explained in Figure~\ref{fig/Sterbenz}: when $x$
and $y$ are sufficiently near one another, cancellation makes the result exactly fit a smaller precision.

\medskip

\fbox{
\begin{minipage}{\fractheo\textwidth}
\begin{theorem}[{{\tt SterbenzApprox2}}]\label{thm:exact-sub}
  Let $x$ and $y$ be $p_1$-digit FPNs. If
  $$
  \frac{y}{1+\beta^{p_2-p_1}} \le x \le \left(1+\beta^{p_2-p_1}\right)\ y,
  $$
  then $x-y$ is a $p_2$-digit FPN.
  This is valid with any different significand sizes $p_1, p_2 \ge 2$, and
  any integer  radix $\beta \ge 2$.
\end{theorem}
\end{minipage}
}

\medskip

The proofs are omitted as they appeared in other publications
\cite{LiBolDau03,BolDau04a}. It is worth mentioning that
Theorem~\ref{thm:exact-sub} do not require $p_1 \ge p_2$ or $p_2 \ge
p_1$.

From now on, all FPNs are binary.
The underlying machine hardware
conforms to IEEE-754 floating point standards~\cite{Ste.87,CodKar.84}. 
This implies that rounding does not introduce a rounding error when the
exact result is a FPN. Unless explicitly stated, the default rounding
mode is {\em round-to-nearest} with ties broken to the even
significand.

\section{About the Algorithm for $z$}\label{sec:alg}

\begin{figure}[hb]
\centerline{\scalebox{.7}{\input{\figdir/kzh.pstex_t}}}
\caption{Algorithm for computing $z$.}
\label{fig/kzh}
\end{figure}
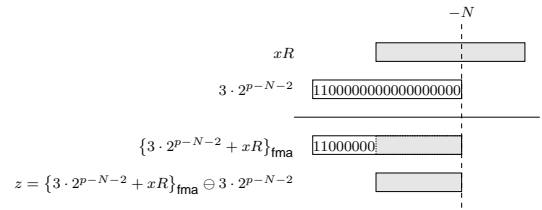

The computation of $z$ can be done efficiently as
\begin{eqnarray}
z & = & \{xR+\sigma\}_{\fma}-\sigma,
\end{eqnarray}
where $\sigma$ is a pre-chosen constant. The technique is adapted from
\cite[Chap.  10]{Mar2K} who used an idea attributed by the author to
C.~Roothaan in his work for HP's vector math library for Itanium.
The explanation is in Figure~\ref{fig/kzh}: here we choose $\sigma=3
\cdot 2^{p-N-2}$ for a $z$ having its last bit at exponent $-N$.

In realizing (\ref{eq:z-extract}), the wanted results are that $z 2^N$ is
an integer, and that $|xR-z| \le 2^{-N-1}$. We may also need that
the precision needed for $z$ is smaller or equal to $p - 2$. Here is
the theorem, verified by Coq.

\medskip

\fbox{
\begin{minipage}{\fractheo\textwidth}
\begin{theorem}[{{\tt arg\_reduct\_exists\_k\_zH}}]\label{theo:zh}
  Assume
  \begin{itemize}
  \item $p > 3$,
  \item $x$ is a $p$-bit FPN,
  \item $R$ is a positive normal $p$-bit FPN,
    \medskip
  \item $z=\left\{3 \cdot 2^{p-N-2}+ x R \right\}_{\fma} \ominus 3  \cdot 2^{p-N-2}$,
    \medskip
  \item $|z| \ge 2^{1-N}$,
  \item $|x R| \le 2^{p-N-2}-2^{-N}$,
  \item $2^{-N}$ is a FPN.
  \end{itemize}
  Then there exists an integer $\ell$ satisfying $2 \le \ell \le p -
  2$ such that
  \begin{itemize}
    \item $|z 2^N|$ is an $\ell$-bit integer greater than $2^{\ell - 1}$, and
    \item $|xR -z| \le 2^{-N-1}$.
  \end{itemize}
\end{theorem}
\end{minipage}
}

\medskip

In short, if $z$ is computed as explained and $x$ is not too big, then
$z$ is  a correct answer, meaning it  fulfills all the requirements
that will be needed in Theorem~\ref{theo:correct1} in the next section.

For Intel's double extended precision, this technique is perfectly
adapted for range reduction with argument between $-2^{63}$ and
$2^{63}$ when $R$ is in the order of $O(1)$.
This argument range coincides with what is in Intel's manual \cite{Int96b}  for {\tt FSIN},
{\tt FCOS}, {\tt FPTAN} and {\tt FSINCOS}. A quick justification is for
$C = 2 \pi$ and modest $N$, say $N = 0$ for an example, $|x R | \lesssim 2^{63} / (2\pi)$
gives
$|x  R | < 2^{62} - 1.$

For the exponential function, any argument larger than $11356$
overflows in the double extended and quad precisions, and $\ell \le p - 2$
is easily satisfied.

\section{Main Results}\label{sec:proof}
\setcounter{equation}{0}

We now present the conditions under which $x - z C_1$ can be
represented exactly by a FPN, and thus it can be computed by $\{x - z
C_1\}_{\fma}$ without error. As in Section~\ref{sec:intro}, $R\approx
1/C$ and $C_1 \approx C$. We suppose that $C > 0$ and $C \neq 2^j$ for
any $j$.

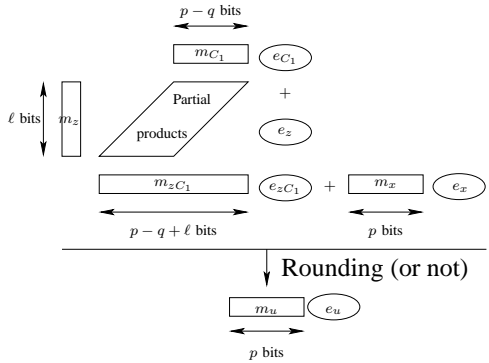
\begin{figure}[h]
\centerline{\scalebox{.7}{\input{\figdir/fmac.pstex_t}}}
\caption{Fused-multiply-add used to create and cancel a higher
  precision virtual number.}
\label{fig/fmac}
\end{figure}

The idea is the use of a fused-multiply-add that may create a higher
precision virtual number that cancels to the working precision.
Figure~\ref{fig/fmac} explains the idea: if $z$ is an $\ell$-bit
integer and the significand of $C_1$ uses $p - q$ bits, it takes up to
$p - q + \ell$ bits to store the significand of $z C_1$.  And as $z
C_1$ and $x$ are near enough, the final result fits into $p$ bits. The
notation $m_X$ stands for the significand of $X$ and $e_X$ its exponent.

We want to give enough hypotheses on the inputs to guarantee that $x -
z C_1$ will be computed without error.

We define the exponent $e_{R}$ of $R$ as the only integer such that
$2^{e_{R}} < R < 2^{e_{R} + 1}$.
We want to set the $q$ least significant bits of  $C_1$ to zero.  Since
$C_1$ should be  as accurate as   possible, we set $C_1  \approx
1/R$ to the nearest FPN with $p -  q$ significant bits. From this, we
deduce that $2^{-e_{R}-1} \le C_1 \le 2^{-e_{R}}$ and that the
distance between $C_1$ and $1/R$ is less than half an ulp (in $p-q$
precision) therefore
$$\left| \frac 1{R} - C_1 \right| \le 2^{- e_{R} - 1 - (p - q)}.$$

We now define $\delta = R C_1-1$, and we deduce a bound on its magnitude from the previous
inequalities
\begin{eqnarray*}
|\delta|  & \le  & 2^{q - p}.
\end{eqnarray*}

Let $z$ be as defined by (\ref{eq:z-extract}) with the conditions on
$z$ and $s$ given there.  We assume for the moment that $z \ne 0$.
Theorem~\ref{thm:exact-sub} can be used if we bound $x/(z C_1)$ and
its reciprocal by $1 + 2^{q - \ell}$. We have the following
equalities:
\begin{eqnarray*}
\frac x{z C_1} & = & \frac {x R}{zR C_1} \\
                 & = & \frac {z+s}{zR C_1} \\
                 & = & \left(1 + \frac sz\right)\frac
                 1{1+\delta}.
\end{eqnarray*}
We recall that $z=k 2^{-N}$ and that $k$ is an integer using $\ell$
bits, and we deduce on the other hand
$$2^{-N + \ell - 1} \le |z| < 2^{-N + \ell}$$
to bound
\begin{eqnarray*}
\left| \frac sz \right| \le 2^{- \ell}.
\end{eqnarray*}

Rewriting the condition of Theorem~\ref{thm:exact-sub} and taking
advantage of preceding results, we arrive at the point
to prove both
\begin{eqnarray}
\label{eq:AB1}\frac{1 + 2^{-\ell}}{1 +|\delta|} & \le & 1 + 2^{q - \ell}, \\[6pt]
\label{eq:AB2}\frac{1 + |\delta|}{1 - 2^{-\ell}} & \le & 1 + 2^{q - \ell}.
\end{eqnarray}

Conditions (\ref{eq:AB1}) and (\ref{eq:AB2}) are checked using functional
analysis on polynomials and homographic functions for any permitted
value of $A = 2^{-\ell}$.  Since $z$ is both a machine number and a
non-zero $\ell$-bit FPN ($1 \le \ell \le p$). From
Section~\ref{sec:alg}, the algorithm used to produce $z$ implies $\ell
\le p - 2$.  We will use a more generic condition:
$$2^{1-p} \le A = 2^{-\ell} \le \frac 12.$$ We will now explain what
are the successive requirements to guarantee that both (\ref{eq:AB1})
and (\ref{eq:AB2}) are fulfilled.

\paragraph{Condition (\ref{eq:AB1})}

We want to guarantee that
$\displaystyle \frac{1 + 2^{-\ell}}{1 + 2^{q - \ell}} \le 1 + |\delta|$.
The homographic function
$$\frac{1 + 2^{-\ell}}{1 + 2^{q - \ell}} = \frac{1 + A}{1 + A 2^q}$$%
we want to bound is maximized at $A = 2^{1-p}$ and it is sufficient to
check if $(1 + 2^{1-p})/(1 + 2^{1-p} 2^q) \le 1 + |\delta|.$ We use the
bound on $|\delta|$ and we introduce $B = 2^q$. We have left to prove
that
$$(1 + 2^{1-p})/(1 + 2^{1-p} B) \le 1 - B 2^{-p}.$$%
This is equivalent to check if the second order polynomial
$2^{1-p} B^2 - B +2 \le 0.$
The inequality is satisfied for $B$ between the two roots
$\displaystyle 2^{p-2} \left(1 \pm \sqrt{1- 2^{4 - p}} \right)$.
Thus it is sufficient  to have $B \ge 4$ for all precisions.

\paragraph{Condition (\ref{eq:AB2})}

We want to guarantee that $1 + |\delta|  \le  (1 + 2^{q - \ell}) (1 -
2^{-\ell})$. We introduce $A$ and $B$ as before, so we have left to prove
$$1 + |\delta|  \le  (1 + AB) (1 - A).$$

We assume   that  $B \ge  4$ from the preceding paragraph. The polynomial
$$(1 + AB)(1 - A) = (1 + 2^{q - \ell})(1 - 2^{-\ell})$$ is minimized
at $A = 2^{1-p}$ and it is sufficient to check if $(1 + |\delta|) \le
(1 + 2^{1-p} B)(1 - 2^{1-p}).$ From the bound on $|\delta|$, we now
have to check if
$$(1 + B 2^{-p}) \le (1 + 2^{1-p} B)(1 - 2^{1-p})$$
which is true for any precision.

This proof is rather long and complex. We therefore verified it in
Coq to be sure there is no mistake. It also gives us more precise
and sharp hypothesis than if we would do that only by pen-and-paper. All
hypotheses have to be clearly written so that the proof can be
checked. There is no easy way to say ``we assume there is no
Underflow'' or ``that the precision is big enough''. This leads to
long theorems (at least longer than what we are used to), but precise
and correct ones:

\medskip

\fbox{
\begin{minipage}{\fractheo\textwidth}
\begin{theorem}[{{\tt Fmac\_arg\_reduct\_correct1}}]\label{theo:correct1}
  Assume
  \begin{itemize}
  \item $p > 3$,
  \item $x$ is a $p$-bit FPN,
  \item $R$ is a positive normal $p$-bit FPN,
  \item $2 \le q < p-1$,
  \item $C_1$ is the $(p-q)$-bit FPN obtained by rounding $1/R$ to
    $p-q$ bits using round-to-nearest mode,
  \item $C_1$ is not exactly a power of 2,
  \item $C_1 \ge 2^{p-q+\max(1,N-1)} \lambda$,
  \item $2 \le \ell \le p-1$,
  \item $|z 2^N|$ is an $\ell$-bit integer greater than $2^{\ell - 1}$,
  \item $|xR -z| \le 2^{-N-1}$,
  \item $q \le \ell$.
  \end{itemize}
  Then $x - z C_1$ is a $p$-bit FPN.
\end{theorem}
\end{minipage}
}

\medskip

In short, if $C_1$ is rounded to the nearest from $1/R$ with $p-q$
bits and $q \ge 2$ and $z$ is not too small, then the {\tt fma} does not
make any round-off error.

Automatic proof checking also prompted us that the exact behavior may be
difficult to obtain for $z=2^{-N}$ and $x$ close to $2^{- N - 1} R$.
This case was excluded in Theorem~\ref{theo:correct1} under the
hypothesis that $2 \le \ell$, but it will be included in the next theorem
which  focuses on $q = 2$ as this situation leads to $C_1$ as close as
possible from $C$ and thus has more practical value.
For completeness and theoretical interest a theorem similar to Theorem~\ref{theo:correct1} but valid for all $2 \le 2 \le p - 1$ is presented in the appendix.

Assume $q=2$ in the rest of this section. When
$z=2^{-N}$, then $x \le 2 C_1 \times 2^{-N}$ as $xR$ is
approximated by $z=2^{-N}$.  We can also deduce that
$$
\frac{C_1 \times  2^{-N}}{1+2^{2-p}}   \le x.
$$
When $C_1 \times 2^{-N}/2 \le x$, Sterbenz's theorem
(Theorem~\ref{thm:exact-sub-classical}) can be applied and $x-C_1
\times 2^{-N}$ is representable.  If not, then
$$
\frac{C_1 \times 2^{-N}}{1+2^{2-p}} \le x < \frac{C_1 \times 2^{-N}}{2}.
$$
Since $C_1$ is a $(p-2)$-bit FPN and not exactly a power of 2 as a $p$-bit FPN, then $C_1$ is at least
4 ulps away from a power of 2. This is because as a $p$-bit FPN,
$C_1$ is worth $2^e\times 1.{\tt bb\cdots b}00$, where at least one of the ${\tt b}$'s must be 1; therefore the $C_1$
that comes closest to a power of $2$ is either  $2^e\times 1.0\cdots 0{\tt 1}00$ or
$2^e\times 1.{\tt 11\cdots 1}00$. Both are 4 ulps away from a power of 2.
This distance and the preceding inequality are enough to guarantee
that the exponent of $x$ is the exponent of $C_1$ minus $N + 1$.
After a few computations, we finish with $x-C_1 \times 2^{-N}$ being a
FPN, regardless of $x$.

A few peculiar cases have been omitted in the sketch of this proof.
Automatic proof checking allows us to trustfully guarantee that these
cases have been all checked in our publicly available proof scripts.
The only surprising condition is presented in this section.  The
other cases are easily generalized from Theorems~\ref{theo:zh} and
\ref{theo:correct1}.  So just by wrapping these two results together,
we can state the following theorem in its full length, verified with
Coq.

\medskip

\fbox{
\begin{minipage}{\fractheo\textwidth}
\begin{theorem}[{{\tt Fmac\_arg\_reduct\_correct3}}]\label{theo:red3}
  Assume
  \begin{itemize}
  \item $p > 3$,
  \item $x$ is a $p$-bit FPN,
  \item $R$ is a positive normal $p$-bit FPN,
  \item $C_1$ is the $(p-2)$-bit FPN obtained by rounding $1/R$ to $p-2$
    bits using round-to-nearest mode,
  \item $C_1$ is not exactly a power of 2,
  \item $C_1 \ge 2^{p+\max(-1,N)} \lambda$,
    \medskip
  \item $z=\left\{3 \cdot 2^{p-N-2}+ x R \right\}_{\fma} \ominus 3  \cdot 2^{p-N-2}$,
    \medskip
  \item $|x R| \le 2^{p-N-2}-2^{-N}$,
  \item $2^{-N}$ is a FPN.
  \end{itemize}
  Then $x - z C_1$ is a $p$-bit FPN.
\end{theorem}
\end{minipage}
}

\medskip

In short, if  $C_1$ is rounded  to  the nearest from  $1/R$ with $p-2$
bits and $z$  is computed as usual, then the  {\tt fma}  does not make  any
round-off error. In  Tables~\ref{table:pi2} and
\ref{table:ln2} we present constants $R$ and $C_1$ for $\pi$ and $\ln(2)$.
These constants are for the exponential and the fast reduction phase of the trigonometric functions \cite{Kah83,Ng92,Mar2K,Mul06}.

The hypotheses may seem numerous and restrictive but they are not. As
$R$ and $C_1$ are pre-computed, the corresponding requirements can be
checked beforehand. Moreover, those requirements are weak: for example
with $0 <= N <= 10$ in double precision, we need $C_1 \ge 2^{-1011} \approx 4.5
10^{-305}$. There is no known elementary function for which $C_1$
ever comes near a power of 2.
The only nontrivial requirement left is the bound on $|xR|$.

\begin{table*}
  \caption{Example of value for $R = \RN(1/C)$, $C_1$ rounded to $p -
    2$ bits, $C_2$ obtained from Algorithm~\ref{algo:C2}, and $C_3$,  for $C = \pi$,
    \newline
    easily leading to $C = 2\pi$ or $C = \pi/2$}
\label{table:pi2}
$$
\begin{array}{l c c c c}
\text{Precision}  & \text{Single}
                  & \text{Double}
                  & \text{Double extended}
                  & \text{Quad} \\[3pt]
R                 & {\scriptstyle 10680707 \cdot 2^{-25}}
                  & {\scriptstyle 5734161139222659 \cdot 2^{-54}}
                  & {\scriptstyle 11743562013128004906 \cdot 2^{-65}}
                  & {\scriptstyle 6611037688290699343682997282138730 \cdot 2^{-114}} \\[3pt]
C_1               & {\scriptstyle 13176796 \cdot 2^{-22}}
                  & {\scriptstyle 7074237752028440 \cdot 2^{-51}}
                  & {\scriptstyle 14488038916154245684 \cdot 2^{-62}}
                  & {\scriptstyle 8156040833015188200833743081374136 \cdot 2^{-111}} \\[3pt]
C_2               & {\scriptstyle -11464520 \cdot 2^{-45}}
                  & {\scriptstyle 4967757600021504 \cdot 2^{-105}}
                  & {\scriptstyle 14179128828124470480 \cdot 2^{-126}}
                  & {\scriptstyle 9351661544631751449372323967920768 \cdot 2^{-226}} \\[3pt]
C_3               & {\scriptstyle -15186280 \cdot 2^{-67}}
                  & {\scriptstyle 7744522442262976 \cdot 2^{-155}}
                  & {\scriptstyle 10700877088903390780 \cdot 2^{-189}}
                  & {\scriptstyle -9186378203702558149401308890796140 \cdot 2^{-334}}
\end{array}
$$
\end{table*}

\begin{table*}
\caption{Example of value for $R = \RN(1/C)$, $C_1$ rounded to $p -
    2$ bits, $C_2$ obtained from Algorithm~\ref{algo:C2}, and $C_3$, for $C = \ln(2)$}
\label{table:ln2}
$$
\begin{array}{l c c c c}
\text{Precision}  & \text{Single}
                  & \text{Double}
                  & \text{Double extended}
                  & \text{Quad} \\[3pt]
R                 & {\scriptstyle 12102203 \cdot 2^{-23}}
                  & {\scriptstyle 6497320848556798 \cdot 2^{-52}}
                  & {\scriptstyle 13306513097844322492 \cdot 2^{-63}}
                  & {\scriptstyle 7490900928631539394323262730195514 \cdot 2^{-112}} \\[3pt]
C_1               & {\scriptstyle 11629080 \cdot 2^{-24}}
                  & {\scriptstyle 6243314768165360 \cdot 2^{-53}}
                  & {\scriptstyle 12786308645202655660 \cdot 2^{-64}}
                  & {\scriptstyle 7198051856247353947080814903691240 \cdot 2^{-113}} \\[3pt]
C_2               & {\scriptstyle -8577792 \cdot 2^{-52}}
                  & {\scriptstyle -7125764960002032 \cdot 2^{-106}}
                  & {\scriptstyle -15596301547560248640 \cdot 2^{-130}}
                  & {\scriptstyle -5381235925004637553074520129202340 \cdot 2^{-224}} \\[3pt]
C_3               & {\scriptstyle -8803384 \cdot 2^{-72}}
                  & {\scriptstyle -7338834209110452 \cdot 2^{-161}}
                  & {\scriptstyle -13766585803531045332 \cdot 2^{-192}}
                  & {\scriptstyle -9437982846677142208552339635087788 \cdot 2^{-338}}
\end{array}
$$
\end{table*}

\section{Getting More Accurate Reduced Arguments}\label{sec:sndstep}
As we pointed out in the introduction in Section~\ref{sec:intro},
sometimes the reduced argument $u=x-z C_1$ is not accurate enough due to the limited precision
in $C_1$ as an approximation to $C$. When this happen another FPN $C_2$ containing the lower bits of the constant $C$ has to be made available and the new reduced argument is now $x-z C_1-z C_2$.
Assume that the conditions of 
Theorem~\ref{theo:red3} hold. In particular $C_1$ has $p-2$ bits in its significand.

The number $x-zC_1 -z C_2$ can be computed exactly \cite{BolMul05} as the sum of two
floats. But here because we know certain conditions on $z$, $C_1$, and $C_2$ as FPNs,
we can do it faster. Inspired by \cite{BolMul05}, we propose the following
Algorithm~\ref{algo:sndstep} to accomplish the task.
It is built upon two known algorithms:
\begin{itemize}
\item Fast2Mult($x$,$y$) that computes the rounded product of $x$ and
$y$ and its error (2 flops) \cite{KarpMarkstein1997}.
\item Fast2Sum($x$,$y$) that computes the rounded sum of $x$ and $y$
and its error (3 flops), under the assumption that  either $x=0$, or
$y=0$, or $|x| \ge |y|$, or  there exist integers $n_x, e_x,
n_y, e_y$ such that $x=n_x 2^{e_x}$ and $y=n_y 2^{e_y}$ and $e_x \ge
e_y$ \cite{DauRidThe01}.
\end{itemize}

\medskip

\begin{algorithm}[Super accurate argument reduction]\label{algo:sndstep}
 \begin{tabular}{ll}
 \dbend & The correctness of this algorithm is only guaranteed \\
 & under the conditions of Theorem~\ref{theo:sndstep}. It does not \\
 & work with any $C_1, C_2$!
 \end{tabular}

\medskip

$$
\begin{array}{rcl}
u   & = & \circ(x - z C_1),\\
v_1 & = & \circ(u - z C_2),\\
(p_1,p_2) & = & \mbox{Fast2Mult}(z, C_2),\\
(t_1,t_2) & = & \mbox{Fast2Sum}(u, - p_1),\\
v_2 & = & \circ(\circ(\circ(t_1-v_1)+t_2)-p_2).
\end{array}
$$
\end{algorithm}

\medskip

\fbox{
\begin{minipage}{\fractheo\textwidth}
\begin{theorem}[{{\tt FArgReduct4.v file}}]\label{theo:sndstep}
  Assume
  \begin{itemize}
  \item $p > 4$,
  \item $x$ is a $p$-bit FPN,
  \item $R$ is a positive normal $p$-bit FPN,
  \item $C_1$ is the $(p-2)$-bit FPN obtained by rounding $1/R$ to $p-2$
    bits using round-to-nearest mode,
  \item $C_1$ it is not exactly a power of 2,
    \medskip
  \item $z=\left\{3 \cdot 2^{p-N-2}+ x R \right\}_{\fma} \ominus 3  \cdot 2^{p-N-2}$,
    \medskip
  \item $|x R| \le 2^{p-N-2}-2^{-N}$,
  \item $2^{-N}$ is a normal $p$-bit FPN,
  \item $C_1 \ge 2^{p+\max(-1,p+N-2)} \lambda$,
  \item $C_2$ is a FPN and an integer multiple of $8\ulp^{\circ 2}
    (C_1)$,
  \item $|C_2| \le 4 \ulp (C_1)$,
  \item $v_1$ and $v_2$ are computed using Algorithm~\ref{algo:sndstep}.
  \end{itemize}
Then Fast2Sum works correctly and we have the mathematical
  equality $v_1+v_2=x - z C_1- z C_2$ (all the computations of the
  last line indeed commit no rounding errors).
\end{theorem}
\end{minipage}
}

\medskip

The first requirements are very similar to the previous ones. The ``no
underflow'' bound on $C_1$ has been raised, but is still easily
achieved by real constants. For a typical $N$ between 0 and 10 used by the existing elementary
math libraries in IEEE double precision, it suffices that $C \ge 10^{-288}$.

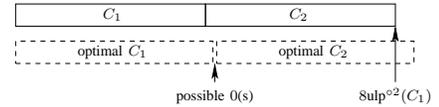
\begin{figure}[h]
\centerline{\scalebox{.7}{\input{\figdir/C1C2.pstex_t}}}
\caption{Respective layouts of our $C_1$ and $C_2$ compared to optimal
values.}
\label{fig/C1C2}
\end{figure}

The most important add-ons are the requirements on $C_2$: it must be
much smaller than $C_1$ (it is near the difference between the
constant $C$ and $C_1$). And $C_2$ must not be ``too precise''.
In fact, $C_1+C_2$
will have $2p-4$ bits as shown in Figure~\ref{fig/C1C2}. If by
chance, there are a lot of zeroes just after $C_1$, we cannot take
advantage of that to get a more precise $C_2$. This is a real
drawback, but it does not happen very often that many zeroes are just
at the inconvenient place.

This algorithm may seem simple but it is a very powerful tool. It is
{\it exact\/} and it
is very {\it fast}: the generic algorithm \cite{BolMul05} costs 20 flops while this one costs only
9 flops! 
More, the result is more usable than expected as it fits in only one
float instead of two in the general case.

As for the computation of $C_2$, the requirements are rather low:
there are several $C_2$ fulfilling them. It may be useful to choose
one of them in order to have the bigger or the smaller $C_2$
possible. Algorithm~\ref{algo:C2} gives one way to compute a
convenient $C_2$.

The idea of the proof for Theorem~\ref{theo:sndstep} is a careful
study of the possible exponents for the involved FPNs. We first prove that
$x$ is an integer multiple of  $2^{-N} \ulp(C_1)$. 
This is done for
whether $z$ is $2^{-N}$ or not to guarantee the correctness of
Fast2Sum.

We then prove that $t_1-v_1$ fits in a FPN. This proof is obtained by
noticing that $t_1$ and $v_1$ are integer multiples of $2^{-N-1}
\ulp^{\circ 2} (C_1)$ and that that $|t_1-v_1| < 2^{p-N-1} \ulp^{\circ
2} (C_1)$.

The next step is about $t_1-v_1+t_2=u-p_1-v_1$ being a FPN. We do it
similarly as all these quantities are also integer multiples of
$2^{-N-1} \ulp^{\circ 2} (C_1)$ and as we easily have that
$|t_1-v_1+t_2| < 2^{p-N-1} \ulp^{\circ 2} (C_1)$.

We finally prove that $t_1-v_1+t_2-p_2=u-z C_2 - v_1$ fits in a FPN.
Its least significant non-zero bit is at most shifted $N$ times down
compared to the least significant non-zero bit of $C_2$. For this
reason, we require that $C_2$ is an integer multiple of $8 \ulp^{\circ
  2} (C_1)$.

This proof needs a careful study of the relationships between the
various floats and their exponent values. The formal proof and its
genericity allowed us a better understanding of the respective layouts
of the FPNs, that is the key of the correctness of
Algorithm~\ref{algo:sndstep}.

\medskip

\begin{algorithm}[Computation of $C_2$]\label{algo:C2}
Let $C$ be the exact constant (for example, $\pi$ or $\ln 2$).
$$
\begin{array}{rcl}
R   & = & \circ_p(1/C),\\
C_1 & = & \circ_{p-2}(1/R), %
\end{array}
$$
and take $C_2$ to be the first many significand bits of $C-C_1$ so that
its least non-zero bit must be greater than or equal to 
$\log_2(\ulp(C_1))-p+4=\log_2\left(8\ulp^{\circ 2} (C_1)\right)$, e.g.,
$$
C_2 =\displaystyle \left\lceil \frac{(C - C_1)}{8\ulp^{\circ 2} (C_1)} \right\rfloor 8\ulp^{\circ 2} (C_1),
$$
where $\lceil\cdot\rfloor$ is one of the round-to-integer operations.
\end{algorithm}

\medskip

This $C_2$ has all the expected properties except that we do not know
for sure if $|C_2| \le 4 \ulp (C_1)$. Note that $C_1$ is not gotten
by directly rounding $C$ but rather
$C_1=\circ\left(1/\circ\left(\frac{1}{C}\right)\right)$.

\medskip

\fbox{
\begin{minipage}{\fractheo\textwidth}
\begin{theorem}[{{\tt gamma2\_le}}]\label{theo:C2}
  Assume
  \begin{itemize}
  \item $p > 3$,
  \item $C$ is a real positive constant,
  \item $R$ is the $p$-bit FPN obtained by rounding $1/C$ to $p$
    bits using round-to-nearest mode,
  \item $R$ is a positive normal $p$-bit FPN,
  \item $C_1$ is the $(p-2)$-bit FPN obtained by rounding $1/R$ to $p-2$
    bits using round-to-nearest mode,
  \item $C_1$ is not exactly a power of 2,
  \item $C_1 \ge 2^{p-1} \lambda$.
  \end{itemize}
Then $|C - C_1| \le 4 \ulp (C_1)$.
\end{theorem}
\end{minipage}
}

\medskip

As $C$ is not too far from $C_1$, we have that $C \le
 2^{p+1} \ulp(C_1) $. We now bound $C-C_1$: \\
$|C - C_1| \le |C- 1/R| + |1/R - C_1| \le
\frac{C}{R} |R-1/C| +|1/R - C_1|$, so 
$|C - C_1| \le
\frac{C}{R} \ulp(R) / 2 + 4 \ulp(C_1) / 2 \le
C 2^{-p-1} + 2 \ulp (C_1)$, hence the result.

\medskip

This means that the formula for $C_2$ given above yields a FPN
fulfilling the requirements of Theorem~\ref{theo:sndstep}.

\section{Conclusions}\label{sec:concl}

We have presented Coq verified theorems that prove the correctness
and effectiveness of a much faster technique based on the commonly
used argument reduction in elementary function computations, on
machines that have hardware support for fused-multiply-add
instructions. The conditions of these theorems are easily met as our
analysis indicates. While we have showed it is not always possible to use
the most accurate parameters under all circumstances, an almost best
possible selection can be used at all times: to zero out the last 2 bits.

We have presented also a very accurate second step argument reduction. We
provide a way to compute $C_2$ which is not the most precise possible,
but is usually 2 bits away from it (and can be rounded as needed by the
programmer). The most interesting part is the possibility to compute
with FPNs the exact error of the second step of the argument reduction
and the fact that this error is exactly representable by only one FPN. It makes the third
step unexpectedly easy as we have a mathematical equality between the
computed FPNs and a very good approximation of $x-z C$ (with a known
error).

Except for the computation of $C_2$, all the rounding used should be
rounding to nearest, ties to even. But our proofs are generic enough
to show that our results still hold when using rounding to nearest,
where cases of ties can be decided in any coherent way
\cite{ReiKnu75}. This includes rounding to nearest, ties away from
zero that is found in the revision of the IEEE-754 standard.

The formal verification forces us to provide many tedious details in
the proofs but gives us a guarantee on our results.  The
proposed theorems are sufficient in the sense that effective
parameters for efficient argument reductions can be obtained without
any difficulty.

Our theorems provides us with sufficient conditions for $x-z C_1$ to be a FPN. This means that $x-z C_1$ could be a FPN  even when one or more of the
conditions fails for some specific values of $C$, $C_1$ and $R$ as
published in the past \cite{StoTan99,Mar2K}.  We may work on
this in the future even though there is only a limited space for
improvement as only the last two bits of $C_1$ can be changed to make
the constant more accurate.

The algorithms proved can be applied to any
floating-point format (IEEE single, double or extended for
example). Intuitively, the correctness of these algorithms should come
as natural. Nevertheless, rigorous proofs are not trivial due to a few special
cases that could  have been easily dismissed by hand-waving proofs.

\bibliographystyle{IEEEtran}
\bibliography{daumas_ref,toto}

\appendix

Theorem~\ref{theo:correct1} can be used for any value of $2 \le q \le
p-1$. In most case, users are interested for the smallest possible
value of $q$ because that will give a more accurate $C_1$ and consequently a more accurate reduced argument. For this reason, we proved Theorem~\ref{theo:red3} for $q=2$.
The following theorem is under the hypothesis that
$$
R C_1 \le 1,
$$
while $2 \le q \le p-1$ still. This add-on is enough to guarantee cases that are left over by
Theorem~\ref{theo:correct1}.

\medskip

\fbox{
\begin{minipage}{\fractheo\textwidth}
\begin{theorem}[{{\tt Fmac\_arg\_reduct\_correct2}}]
  Assume
  \begin{itemize}
  \item $p > 3$,
  \item $2 \le q < p-1$,
  \item $x$ is $p$-bit FPN,
  \item $R$ is a positive normal $p$-bit FPN,
  \item $C_1$ is  the $(p-q)$-bit  FPN obtained by  rounding
    $1/R$ to $p-q$ bits using round-to-nearest mode,
  \item $C_1$ is not exactly a power of 2,
  \item $C_1 \ge 2^{p-q+\max(1,N-1)} \lambda$,
    \medskip
  \item $z=\left\{3 \cdot 2^{p-N-2}+ x R \right\}_{\fma} \ominus 3  \cdot 2^{p-N-2}$,
    \medskip
  \item $2^{-N}$ is a FPN,
  \item $|x R| \le 2^{p-N-2}-2^{-N}$,
  \item $R C_1 \le 1$.
  \end{itemize}
  Then $x - z C_1$ is a $p$-bit FPN.
\end{theorem}
\end{minipage}
}

\medskip

We essentially need to consider how to make $R$ and $C_1$ satisfy
this new constraint.  Since there is no strict connection between
$R$, $C_1$ on one hand and $C$ on the other hand, we can either use
$R$ to be the correctly rounded FPN nearest $1/C$ or we may
alternatively add or subtract one or a few ulps so that
the additional inequality is met.

\end{document}

%% file: cody.pstex_t
\begin{picture}(0,0)%
\includegraphics{\figdir/cody.pstex}%
\end{picture}%
\setlength{\unitlength}{2930sp}%
\begingroup\makeatletter\ifx\SetFigFont\undefined%
\gdef\SetFigFont#1#2#3#4#5{%
  \reset@font\fontsize{#1}{#2pt}%
  \fontfamily{#3}\fontseries{#4}\fontshape{#5}%
  \selectfont}%
\fi\endgroup%
\begin{picture}(5168,3670)(-1420,-3044)
\put(226,434){\makebox(0,0)[b]{\smash{{\SetFigFont{8}{9.6}{\familydefault}{\mddefault}{\updefault}{\color[rgb]{0,0,0}$x$}%
}}}}
\put(676,-241){\makebox(0,0)[b]{\smash{{\SetFigFont{8}{9.6}{\familydefault}{\mddefault}{\updefault}{\color[rgb]{0,0,0}$z$}%
}}}}
\put(-449,-241){\makebox(0,0)[b]{\smash{{\SetFigFont{8}{9.6}{\familydefault}{\mddefault}{\updefault}{\color[rgb]{0,0,0}$C_1$}%
}}}}
\put(  1,-691){\makebox(0,0)[b]{\smash{{\SetFigFont{8}{9.6}{\familydefault}{\mddefault}{\updefault}{\color[rgb]{0,0,0}$z  \times C_1$ is exact}%
}}}}
\put(676,-1366){\makebox(0,0)[b]{\smash{{\SetFigFont{8}{9.6}{\familydefault}{\mddefault}{\updefault}{\color[rgb]{0,0,0}$x-z  \times C_1$ is exact}%
}}}}
\put(1576,-1816){\makebox(0,0)[b]{\smash{{\SetFigFont{8}{9.6}{\familydefault}{\mddefault}{\updefault}{\color[rgb]{0,0,0}$\circ(z \times C_2)$}%
}}}}
\put(1081,-2491){\makebox(0,0)[b]{\smash{{\SetFigFont{8}{9.6}{\familydefault}{\mddefault}{\updefault}{\color[rgb]{0,0,0}$\circ(x-z \times C_1 - \circ(z \times C_2))$}%
}}}}
\put(1081,-2986){\makebox(0,0)[b]{\smash{{\SetFigFont{8}{9.6}{\familydefault}{\mddefault}{\updefault}{\color[rgb]{0,0,0}final precision}%
}}}}
\put(2431,-2491){\makebox(0,0)[lb]{\smash{{\SetFigFont{8}{9.6}{\familydefault}{\mddefault}{\updefault}{\color[rgb]{0,0,0}round-off error}%
}}}}
\put(-359,-1366){\makebox(0,0)[rb]{\smash{{\SetFigFont{8}{9.6}{\familydefault}{\mddefault}{\updefault}{\color[rgb]{0,0,0}cancellation}%
}}}}
\end{picture}%

%% file: cody2.pstex_t
\begin{picture}(0,0)%
\includegraphics{\figdir/cody2.pstex}%
\end{picture}%
\setlength{\unitlength}{2930sp}%
\begingroup\makeatletter\ifx\SetFigFont\undefined%
\gdef\SetFigFont#1#2#3#4#5{%
  \reset@font\fontsize{#1}{#2pt}%
  \fontfamily{#3}\fontseries{#4}\fontshape{#5}%
  \selectfont}%
\fi\endgroup%
\begin{picture}(4804,3670)(-1191,-3044)
\put(226,434){\makebox(0,0)[b]{\smash{{\SetFigFont{8}{9.6}{\familydefault}{\mddefault}{\updefault}{\color[rgb]{0,0,0}$x$}%
}}}}
\put(  1,-691){\makebox(0,0)[b]{\smash{{\SetFigFont{8}{9.6}{\familydefault}{\mddefault}{\updefault}{\color[rgb]{0,0,0}$z  \times C_1$ is exact}%
}}}}
\put(-899,-241){\makebox(0,0)[b]{\smash{{\SetFigFont{8}{9.6}{\familydefault}{\mddefault}{\updefault}{\color[rgb]{0,0,0}$C_1$}%
}}}}
\put(226,-241){\makebox(0,0)[b]{\smash{{\SetFigFont{8}{9.6}{\familydefault}{\mddefault}{\updefault}{\color[rgb]{0,0,0}$z$}%
}}}}
\put(451,-1366){\makebox(0,0)[b]{\smash{{\SetFigFont{8}{9.6}{\familydefault}{\mddefault}{\updefault}{\color[rgb]{0,0,0}$x-z  \times C_1$ is exact}%
}}}}
\put(1576,-2986){\makebox(0,0)[b]{\smash{{\SetFigFont{8}{9.6}{\familydefault}{\mddefault}{\updefault}{\color[rgb]{0,0,0}final precision}%
}}}}
\put(676,-1816){\makebox(0,0)[b]{\smash{{\SetFigFont{8}{9.6}{\familydefault}{\mddefault}{\updefault}{\color[rgb]{0,0,0}$\circ(z \times C_2)$}%
}}}}
\put(1576,-2491){\makebox(0,0)[b]{\smash{{\SetFigFont{8}{9.6}{\familydefault}{\mddefault}{\updefault}{\color[rgb]{0,0,0}$\circ(x-z \times C_1 - \circ(z \times C_2))$}%
}}}}
\put(2116,-1816){\makebox(0,0)[lb]{\smash{{\SetFigFont{8}{9.6}{\familydefault}{\mddefault}{\updefault}{\color[rgb]{0,0,0}round-off error}%
}}}}
\end{picture}%

%% file: enh.pstex_t
\begin{picture}(0,0)%
\includegraphics{\figdir/enh.pstex}%
\end{picture}%
\setlength{\unitlength}{2930sp}%
\begingroup\makeatletter\ifx\SetFigFont\undefined%
\gdef\SetFigFont#1#2#3#4#5{%
  \reset@font\fontsize{#1}{#2pt}%
  \fontfamily{#3}\fontseries{#4}\fontshape{#5}%
  \selectfont}%
\fi\endgroup%
\begin{picture}(7179,3670)(-1136,-3044)
\put(226,434){\makebox(0,0)[b]{\smash{{\SetFigFont{8}{9.6}{\familydefault}{\mddefault}{\updefault}{\color[rgb]{0,0,0}$x$}%
}}}}
\put(2431,-241){\makebox(0,0)[b]{\smash{{\SetFigFont{8}{9.6}{\familydefault}{\mddefault}{\updefault}{\color[rgb]{0,0,0}$z$}%
}}}}
\put(226,-241){\makebox(0,0)[b]{\smash{{\SetFigFont{8}{9.6}{\familydefault}{\mddefault}{\updefault}{\color[rgb]{0,0,0}$C_1$}%
}}}}
\put(1126,-691){\makebox(0,0)[b]{\smash{{\SetFigFont{8}{9.6}{\familydefault}{\mddefault}{\updefault}{\color[rgb]{0,0,0}$z  \times C_1$}%
}}}}
\put(2251,-1366){\makebox(0,0)[b]{\smash{{\SetFigFont{8}{9.6}{\familydefault}{\mddefault}{\updefault}{\color[rgb]{0,0,0}$x-z  \times C_1$ is exact}%
}}}}
\put(  1,-1366){\makebox(0,0)[b]{\smash{{\SetFigFont{8}{9.6}{\familydefault}{\mddefault}{\updefault}{\color[rgb]{0,0,0}cancellation}%
}}}}
\put(2476,-2491){\makebox(0,0)[b]{\smash{{\SetFigFont{8}{9.6}{\familydefault}{\mddefault}{\updefault}{\color[rgb]{0,0,0}$\circ(x-z \times C_1 - z \times C_2)$}%
}}}}
\put(2431,-2986){\makebox(0,0)[b]{\smash{{\SetFigFont{8}{9.6}{\familydefault}{\mddefault}{\updefault}{\color[rgb]{0,0,0}final precision}%
}}}}
\put(3826,-1816){\makebox(0,0)[b]{\smash{{\SetFigFont{8}{9.6}{\familydefault}{\mddefault}{\updefault}{\color[rgb]{0,0,0}$z \times C_2$}%
}}}}
\put(3871,-2491){\makebox(0,0)[lb]{\smash{{\SetFigFont{8}{9.6}{\familydefault}{\mddefault}{\updefault}{\color[rgb]{0,0,0}round-off error}%
}}}}
\end{picture}%

%% file: Sterbenz.pstex_t
\begin{picture}(0,0)%
\includegraphics{\figdir/Sterbenz.pstex}%
\end{picture}%
\setlength{\unitlength}{2930sp}%
\begingroup\makeatletter\ifx\SetFigFont\undefined%
\gdef\SetFigFont#1#2#3#4#5{%
  \reset@font\fontsize{#1}{#2pt}%
  \fontfamily{#3}\fontseries{#4}\fontshape{#5}%
  \selectfont}%
\fi\endgroup%
\begin{picture}(2049,2059)(2239,-1694)
\put(3376,209){\makebox(0,0)[b]{\smash{{\SetFigFont{8}{9.6}{\familydefault}{\mddefault}{\updefault}{\color[rgb]{0,0,0}$p_1$ digits}%
}}}}
\put(3331,-241){\makebox(0,0)[b]{\smash{{\SetFigFont{8}{9.6}{\familydefault}{\mddefault}{\updefault}{\color[rgb]{0,0,0}$x$}%
}}}}
\put(2431,-601){\makebox(0,0)[b]{\smash{{\SetFigFont{8}{9.6}{\familydefault}{\mddefault}{\updefault}{\color[rgb]{0,0,0}$-$}%
}}}}
\put(3331,-601){\makebox(0,0)[b]{\smash{{\SetFigFont{8}{9.6}{\familydefault}{\mddefault}{\updefault}{\color[rgb]{0,0,0}$y$}%
}}}}
\put(3691,-1141){\makebox(0,0)[b]{\smash{{\SetFigFont{8}{9.6}{\familydefault}{\mddefault}{\updefault}{\color[rgb]{0,0,0}$x-y$}%
}}}}
\put(3691,-1636){\makebox(0,0)[b]{\smash{{\SetFigFont{8}{9.6}{\familydefault}{\mddefault}{\updefault}{\color[rgb]{0,0,0}$p_2$ digits}%
}}}}
\end{picture}%

%% file: kzh.pstex_t
\begin{picture}(0,0)%
\includegraphics{\figdir/kzh.pstex}%
\end{picture}%
\setlength{\unitlength}{2930sp}%
\begingroup\makeatletter\ifx\SetFigFont\undefined%
\gdef\SetFigFont#1#2#3#4#5{%
  \reset@font\fontsize{#1}{#2pt}%
  \fontfamily{#3}\fontseries{#4}\fontshape{#5}%
  \selectfont}%
\fi\endgroup%
\begin{picture}(5919,2499)(-4961,-1423)
\put(  1,929){\makebox(0,0)[b]{\smash{{\SetFigFont{8}{9.6}{\familydefault}{\mddefault}{\updefault}{\color[rgb]{0,0,0}$-N$}%
}}}}
\put(-2024,434){\makebox(0,0)[rb]{\smash{{\SetFigFont{8}{9.6}{\familydefault}{\mddefault}{\updefault}{\color[rgb]{0,0,0}$xR$}%
}}}}
\put(-2024,-16){\makebox(0,0)[rb]{\smash{{\SetFigFont{8}{9.6}{\familydefault}{\mddefault}{\updefault}{\color[rgb]{0,0,0}$3 \cdot 2^{p-N-2}$}%
}}}}
\put(-2024,-691){\makebox(0,0)[rb]{\smash{{\SetFigFont{8}{9.6}{\familydefault}{\mddefault}{\updefault}{\color[rgb]{0,0,0}$\left\{3 \cdot 2^{p-N-2}+ x R \right\}_{\fma}$}%
}}}}
\put(-2024,-1141){\makebox(0,0)[rb]{\smash{{\SetFigFont{8}{9.6}{\familydefault}{\mddefault}{\updefault}{\color[rgb]{0,0,0}$z=\left\{3 \cdot 2^{p-N-2}+ x R \right\}_{\fma} \ominus 3  \cdot 2^{p-N-2}$}%
}}}}
\put(-1799,-16){\makebox(0,0)[lb]{\smash{{\SetFigFont{8}{9.6}{\familydefault}{\mddefault}{\updefault}{\color[rgb]{0,0,0}$1100000000000000000$}%
}}}}
\put(-1799,-691){\makebox(0,0)[lb]{\smash{{\SetFigFont{8}{9.6}{\familydefault}{\mddefault}{\updefault}{\color[rgb]{0,0,0}$11000000$}%
}}}}
\end{picture}%

%% file: fmac.pstex_t
\begin{picture}(0,0)%
\includegraphics{\figdir/fmac.pstex}%
\end{picture}%
\setlength{\unitlength}{2930sp}%
\begingroup\makeatletter\ifx\SetFigFont\undefined%
\gdef\SetFigFont#1#2#3#4#5{%
  \reset@font\fontsize{#1}{#2pt}%
  \fontfamily{#3}\fontseries{#4}\fontshape{#5}%
  \selectfont}%
\fi\endgroup%
\begin{picture}(5592,4371)(-224,-3454)
\put(2881,-2401){\makebox(0,0)[lb]{\smash{{\SetFigFont{14}{16.8}{\familydefault}{\mddefault}{\updefault}{\color[rgb]{0,0,0}Rounding (or not)}%
}}}}
\put(5041,-1366){\makebox(0,0)[b]{\smash{{\SetFigFont{8}{9.6}{\familydefault}{\mddefault}{\updefault}{\color[rgb]{0,0,0}$e_x$}%
}}}}
\put(4141,-1321){\makebox(0,0)[b]{\smash{{\SetFigFont{8}{9.6}{\familydefault}{\mddefault}{\updefault}{\color[rgb]{0,0,0}$m_x$}%
}}}}
\put(4141,-1906){\makebox(0,0)[b]{\smash{{\SetFigFont{8}{9.6}{\familydefault}{\mddefault}{\updefault}{\color[rgb]{0,0,0}$p$ bits}%
}}}}
\put(2926,209){\makebox(0,0)[b]{\smash{{\SetFigFont{8}{9.6}{\familydefault}{\mddefault}{\updefault}{\color[rgb]{0,0,0}$e_{C_1}$}%
}}}}
\put(2926,-691){\makebox(0,0)[b]{\smash{{\SetFigFont{8}{9.6}{\familydefault}{\mddefault}{\updefault}{\color[rgb]{0,0,0}$e_z$}%
}}}}
\put(2926,-241){\makebox(0,0)[b]{\smash{{\SetFigFont{8}{9.6}{\familydefault}{\mddefault}{\updefault}{\color[rgb]{0,0,0}$+$}%
}}}}
\put(2926,-1366){\makebox(0,0)[b]{\smash{{\SetFigFont{8}{9.6}{\familydefault}{\mddefault}{\updefault}{\color[rgb]{0,0,0}$e_{z C_1}$}%
}}}}
\put(2701,-2851){\makebox(0,0)[b]{\smash{{\SetFigFont{8}{9.6}{\familydefault}{\mddefault}{\updefault}{\color[rgb]{0,0,0}$m_u$}%
}}}}
\put(2026,254){\makebox(0,0)[b]{\smash{{\SetFigFont{8}{9.6}{\familydefault}{\mddefault}{\updefault}{\color[rgb]{0,0,0}$m_{C_1}$}%
}}}}
\put(1576,-1321){\makebox(0,0)[b]{\smash{{\SetFigFont{8}{9.6}{\familydefault}{\mddefault}{\updefault}{\color[rgb]{0,0,0}$m_{z C_1}$}%
}}}}
\put(2026,749){\makebox(0,0)[b]{\smash{{\SetFigFont{8}{9.6}{\familydefault}{\mddefault}{\updefault}{\color[rgb]{0,0,0}$p-q$ bits}%
}}}}
\put(2701,-3391){\makebox(0,0)[b]{\smash{{\SetFigFont{8}{9.6}{\familydefault}{\mddefault}{\updefault}{\color[rgb]{0,0,0}$p$ bits}%
}}}}
\put(1576,-1906){\makebox(0,0)[b]{\smash{{\SetFigFont{8}{9.6}{\familydefault}{\mddefault}{\updefault}{\color[rgb]{0,0,0}$p-q+\ell$ bits}%
}}}}
\put(3511,-2851){\makebox(0,0)[b]{\smash{{\SetFigFont{8}{9.6}{\familydefault}{\mddefault}{\updefault}{\color[rgb]{0,0,0}$e_u$}%
}}}}
\put(1801,-331){\makebox(0,0)[b]{\smash{{\SetFigFont{8}{9.6}{\familydefault}{\mddefault}{\updefault}{\color[rgb]{0,0,0}Partial}%
}}}}
\put(1441,-736){\makebox(0,0)[b]{\smash{{\SetFigFont{8}{9.6}{\familydefault}{\mddefault}{\updefault}{\color[rgb]{0,0,0}products}%
}}}}
\put(-224,-556){\makebox(0,0)[b]{\smash{{\SetFigFont{8}{9.6}{\familydefault}{\mddefault}{\updefault}{\color[rgb]{0,0,0}$\ell$ bits}%
}}}}
\put(3466,-1366){\makebox(0,0)[b]{\smash{{\SetFigFont{8}{9.6}{\familydefault}{\mddefault}{\updefault}{\color[rgb]{0,0,0}$+$}%
}}}}
\put(316,-556){\makebox(0,0)[b]{\smash{{\SetFigFont{8}{9.6}{\familydefault}{\mddefault}{\updefault}{\color[rgb]{0,0,0}$m_z$}%
}}}}
\end{picture}%

%% file: C1C2.pstex_t
\begin{picture}(0,0)%
\includegraphics{\figdir/C1C2.pstex}%
\end{picture}%
\setlength{\unitlength}{2930sp}%
\begingroup\makeatletter\ifx\SetFigFont\undefined%
\gdef\SetFigFont#1#2#3#4#5{%
  \reset@font\fontsize{#1}{#2pt}%
  \fontfamily{#3}\fontseries{#4}\fontshape{#5}%
  \selectfont}%
\fi\endgroup%
\begin{picture}(4839,1245)(-1136,-619)
\put(3466,-556){\makebox(0,0)[b]{\smash{{\SetFigFont{8}{9.6}{\familydefault}{\mddefault}{\updefault}{\color[rgb]{0,0,0}$8\ulp^{\circ 2}(C_1)$}%
}}}}
\put( 46,-16){\makebox(0,0)[b]{\smash{{\SetFigFont{8}{9.6}{\familydefault}{\mddefault}{\updefault}{\color[rgb]{0,0,0}optimal $C_1$}%
}}}}
\put( 46,434){\makebox(0,0)[b]{\smash{{\SetFigFont{8}{9.6}{\familydefault}{\mddefault}{\updefault}{\color[rgb]{0,0,0}$C_1$}%
}}}}
\put(2296,434){\makebox(0,0)[b]{\smash{{\SetFigFont{8}{9.6}{\familydefault}{\mddefault}{\updefault}{\color[rgb]{0,0,0}$C_2$}%
}}}}
\put(2476,-16){\makebox(0,0)[b]{\smash{{\SetFigFont{8}{9.6}{\familydefault}{\mddefault}{\updefault}{\color[rgb]{0,0,0}optimal $C_2$}%
}}}}
\put(1284,-556){\makebox(0,0)[b]{\smash{{\SetFigFont{8}{9.6}{\familydefault}{\mddefault}{\updefault}{\color[rgb]{0,0,0}possible $0$(s)}%
}}}}
\end{picture}%